\documentclass[aps,prl,nofootinbib,preprintnumbers,amsmath,amssymb,latexsym,array,enumerate,letter,twocolumn]{revtex4}
\usepackage{amssymb}
\usepackage{amsmath}
\usepackage{epsfig}
\usepackage{hyperref}
\usepackage{breakurl}
\usepackage{xcolor}
\makeatletter
\def\simgt{\mathrel{\lower2.5pt\vbox{\lineskip=0pt\baselineskip=0pt
           \hbox{$>$}\hbox{$\sim$}}}}
\def\simlt{\mathrel{\lower2.5pt\vbox{\lineskip=0pt\baselineskip=0pt
           \hbox{$<$}\hbox{$\sim$}}}}
\makeatother

\def\mysection#1{{\bf #1.} }

\newcommand{\be}{\begin{equation}}
\newcommand{\ee}{\end{equation}}
\newcommand{\bea}{\begin{eqnarray}}
\newcommand{\eea}{\end{eqnarray}}
\newcommand{\beq}{\begin{eqnarray}}
\newcommand{\eeq}{\end{eqnarray}}

\def\mysection#1{{\bf #1.} }

\def\lsim{\mathrel{\rlap{\lower4pt\hbox{\hskip1pt$\sim$}}
     \raise1pt\hbox{$<$}}}         %less than or approx. symbol
\def\gsim{\mathrel{\rlap{\lower4pt\hbox{\hskip1pt$\sim$}}
     \raise1pt\hbox{$>$}}}         %greater than or approx. symbol

\begin{document}

\widetext
%\leftline{PP-017-28} 

\title{Primordial Anisotropies in the Gravitational Wave Background \\ 
from Cosmological Phase Transitions }

\author{Michael  Geller, Anson   Hook, Raman  Sundrum, Yuhsin  Tsai}

\affiliation{~ \\
Maryland Center for Fundamental Physics, Department of Physics,
University of Maryland, College Park, MD 20742, USA}

\begin{abstract}
Phase transitions in the early universe can readily create an observable stochastic gravitational wave background.
We show that such a background necessarily contains anisotropies analogous to those of the cosmic microwave background (CMB) of photons, and   
that these too may be within reach of proposed gravitational wave detectors. Correlations within the gravitational wave anisotropies and their cross-correlations with the CMB
can provide new insights into the mechanism underlying primordial fluctuations, such as multi-field inflation, as well as reveal the existence of non-standard ``hidden sectors" of particle physics in earlier eras.

\end{abstract}

\maketitle

With the recent observation of gravitational waves (GW) by LIGO-VIRGO~\cite{TheLIGOScientific:2016wyq,TheLIGOScientific:2017qsa} we have entered a new era of astronomy, which will illuminate the most mysterious objects in the sky such as black holes and neutron stars.  Remarkably, with future improvements in GW detectors we will also enter a new era of observational {\it cosmology}. The early universe can readily contain a variety of GW sources - inflationary fluctuations~\cite{Starobinsky:1979ty}, cosmological particle production~\cite{Cook:2011hg,Senatore:2011sp}, (p)reheating~\cite{Khlebnikov:1997di}, phase transitions (PT)~\cite{Steinhardt:1981ct,Kosowsky:1992rz,Kamionkowski:1993fg,Nicolis:2003tg,Grojean:2006bp,Randall:2006py,Caprini:2015zlo,AmaroSeoane:2012km,Schwaller:2015tja}, and cosmic strings~\cite{Vachaspati:1984gt,Sakellariadou:1990ne}. 

%The case for phase transitions (PT) is particularly well motivated by theoretical and observational considerations. 
Many extensions of the Standard Model (SM) of particle physics, whether ambitious paradigms such as supersymmetry or particle compositeness, or relatively modest extensions (such as the addition of just one gauge singlet scalar field), can undergo strong first order PT at high temperatures of order the electroweak scale or higher, $T\gsim$ TeV. The PT proceed through the nucleation and dynamics of large bubbles of the low temperature phase, acting as a strong source of long wavelength GW. (See Ref. \cite{Caprini:2015zlo} for a review.) This GW background is analogous to the cosmic microwave background (CMB) of 3K photons. Just as CMB radiation is emitted from the surface of last scattering, the GW are effectively emitted from a distant surface at the periphery of our universe at the time of the PT. These GW then travel to us undergoing a significant cosmological redshift en route, thereby obtaining frequencies, $\omega_{\rm GW} \sim \text{mHz - Hz}$, and signal strength that can be detected at proposed GW observatories such as LISA~\cite{2017arXiv170200786A}, BBO~\cite{Harry:2006fi}, MAGIS~\cite{Graham:2017pmn}, DECIGO~\cite{Kawamura:2011zz}, ALIA~\cite{Gong:2014mca}, or even LIGO \cite{Dev:2016feu}.
%\footnote{Higher temperature PT are certainly possible, perhaps in the LIGO frequency range. See for example \cite{Dev:2016feu}. Grand Unification would also imply a high temperature PT, but 
 %with GW signals well outside the reach of proposed detectors \cite{Kosowsky:1992rz}. }. 
 While the fates of individual bubbles formed during the PT are essentially random, the resulting GW would be seen today as a diffuse background arriving from all directions, coarse-grained over an extremely large number of bubbles. The detailed frequency spectrum of this stochastic GW background would reflect the physics of the PT, during a cosmological era otherwise difficult to access. Furthermore, particle collider experiments, such as the CERN LHC or beyond, may provide a complementary view of the associated particle physics (for example, see Ref.~\cite{Arkani-Hamed:2104388}). 

Stochastic GW  generated by cosmological PT will appear as an approximately
 isotropic background with average energy density $\bar{\rho}_{\text{GW}}$.  
In this paper we will argue that there necessarily are also anisotropies in this background, $\rho_{\rm GW} = \bar{\rho}_{\text{GW}} + \delta \rho_{\rm GW}$,
again analogous to the CMB, providing  a unique window onto 
the physics generating primordial inhomogeneities, plausibly during an inflationary era  well before the PT itself. (See Ref.~\cite{Baumann:2009ds} for a review of inflation.) 
% We will show that while the GW background anisotropies might very naturally be aligned with the CMB anisotropies, they may also be completely independent, depending on key features of the inflationary dynamics. 
We will show that  such anisotropies may be accessible with sensitive directional detection at the proposed GW observatories.
  In this way while the GW frequency spectrum can teach us about the physics of the PTs at multi-TeV scales, the anisotropies can teach us about physics at vastly higher energy 
scales.%\footnote{Famously, anisotropic gravitational waves can be produced {\it during} high-scale inflation, expanded to very large ``super-horizon" wavelengths and imprinted on the (polarized) CMB~\cite{Zaldarriaga:1996xe,Kamionkowski:1996ks} or re-enter the horizon directly as a very weak GW background~\cite{Smith:2005mm}. This is quite distinct from the physics pursued here.}

The key observable is given by the differential energy density of GW arriving from an infinitesimal solid angle of the sky:
\begin{equation}
d\rho_{\text{GW}}=\rho_{\text{GW}}(\theta,\phi) \,\sin\theta \,d\theta\,d\phi.
\end{equation}
From this we can compute two-point correlations,
\begin{equation}
C^{GW}(\theta) \equiv \frac{\left<\rho_{\rm GW} (1)\, \rho_{\rm GW} (2) \right>_{ \theta}}{\bar{\rho}^2_{\rm GW}},
\end{equation}
where we are averaging over all pairs of points on the sphere,  $1,2$, separated by a fixed angle $\theta$. It is standard to expand anisotropies in 
Legendre polynomials,
\begin{equation}\label{eq:ps}
C^{GW}(\theta) = \frac{1}{4\pi} \sum_{l} {\left(2l+1\right)  \,C_{\ell}^{\text{GW}}P_l\left(\cos\theta\right)}.
\end{equation}
As we will show, the GW background and the CMB can share the same primordial source of fluctuations or have quite different origins, and this will be visible in the cross-correlations with the CMB, 
\begin{equation}
C^{cross}(\theta) \equiv \frac{\left<\rho_{\rm GW} (1) \,\rho_{\rm CMB} (2) \right>_{ \theta}}{\bar{\rho}_{\rm GW}\, \bar{\rho}_{\rm CMB}}.
\end{equation}
%Higher-point GW correlators may also be observable. 

%  It may also be possible for higher-point correlators, such as  $\left<\rho_{\rm GW} (1) \rho_{\rm GW} (2)   \rho_{\rm GW}(3)\right>$, to be observable, reflecting interactions within the inflationary dynamics. 

Anisotropic GW backgrounds have been considered earlier.   Refs.~\cite{Cutler:2009qv,Cusin:2017fwz,Cusin:2017mjm} have proposed anisotropic signals due to astrophysical sources, such as white dwarf mergers. Such anisotropies reflect both the inhomogeneous distribution of sources as well the gravitational lensing of the GW by (dark) matter as they propagate to us. This would yield an independent measure of the matter power spectrum.  Refs.~\cite{Bethke:2013aba,Bethke:2013vca} have studied inflationary preheating as a source of very high frequency GW $\omega_{\rm GW} \sim$MHz-GHz, although this is currently  challenging for GW detection. Refs.
\cite{Olmez:2011cg,Contaldi:2016koz,Cusin:2017fwz,Jenkins:2018nty} developed analytic frameworks for characterizing the anisotropies in GW. Ref.~\cite{Cusin:2017fwz,Cusin:2018rsq} applied their formalism to the case of astrophysical mergers, while Ref.~\cite{Jenkins:2018nty} generalized the framework to include cosmic string networks as the GW source. 

The present paper makes four main points. (i) First-order PTs in multi-TeV extensions of the SM  are a robust and plausible  source of {\it anisotropic} GW. 
 (ii) The anisotropies are almost completely primordial in nature, directly reflecting the era of inflation. (iii) The GW anisotropies can exhibit a variety of behaviors, including $(\delta \rho/\rho)_{\rm GW} \gg (\delta \rho/\rho)_{\rm CMB}$ and/or varying degrees of cross-correlation with the CMB, sensitive to the nature of inflation and reheating. 
(iv) Current cosmological data constrain the GW background,  but nevertheless there is considerable scope for detecting a range of possible anisotropies at the next generation of detectors, including LISA. 

There are three contributing processes to the GW signal from PT: bubble wall collisions, sound waves, and magnetohydrodynamic (MHD) turbulence in the plasma. The latter two mechanisms depend on more details of the plasma and model-specifics and are topics of active research. Although simulation results suggest larger signals are produced from sound waves and MHD turbulence~\cite{Hindmarsh:2015qta,Caprini:2015zlo}, here we focus on just the signal from the envelope approximation of the bubble wall collisions~\cite{Kosowsky:1991ua}, which is currently better understood. Since we will in general be considering challengingly small GW backgrounds and anisotropies, this approach is quite conservative.  

The energy in GW can conveniently be expressed in terms of the energy in CMB photons today, $\rho_{\gamma}$,
\begin{equation}
\rho_{\rm GW}^{today} = 0.06 \frac{\rho_{PT}^2}{\rho_{total}^2} 
%\left(H_{PT} \Delta t_{PT} \right)^2 
\left(\frac{H_{PT}}{\beta}\right)^2\rho_{\gamma},
\label{eq:rhoGW}
\end{equation}
as reviewed in Ref.~\cite{Konstandin:2017sat,Cutting:2018tjt}.\footnote{We assume the bubble wall moves at the speed of light, and all of the latent heat goes into GW.} Here, $\rho_{PT}$ is the energy density in the sector undergoing the PT, 
$H_{PT} \sim \sqrt{G_N \rho_{total}}$ is the Hubble expansion rate at the time of the PT, and 
  $\beta \equiv - \dot{S}_{bounce}$ at PT, where $S_{bounce}$ is the tunneling bounce action, with 
\begin{equation}\label{eq:HT}
\frac{\beta}{H_{PT}}={\rm few} - {\cal O}(100).
\end{equation}
This range is spanned by  models in the literature with strongly first-order PT~\cite{Caprini:2015zlo}. 
The larger values in this range are natural but the smaller values can arise with modest tuning of microphysical couplings. 
We have assumed that the PT remnants consist of just SM $+$ dark matter (DM) $+$ GW, as motivated in extensions of the SM related to the electroweak hierarchy problem (reviewed in Ref.~\cite{Csaki:2016kln}) or electroweak scale baryogenesis (reviewed in Ref.~\cite{Morrissey:2012db}). 
Clearly the largest signal would arise if the entire contents of the universe undergoes the PT, $\rho_{total} = \rho_{PT}$.

The GW frequencies are also redshifted \cite{Konstandin:2017sat,Cutting:2018tjt}, with peak frequency
\begin{eqnarray}\label{eq:ftoday}
\omega^{today}_{\rm GW} &=& 0.03\,{\rm mHz}~
\left(\frac{\beta}{H_{PT}}\right) \left(\frac{T_{PT}}{\rm TeV}\right).
\end{eqnarray}
For 
%where we get ballpark detectable frequencies for 
PTs at temperatures $T_{PT} \gsim10$ TeV, most of the integrated GW signal can fit in the $\sim 1 - 10~ {\rm mHz}$ frequency range covered by the LISA detector, and above its  expected sensitivity there\footnote{We assume the LISA proposal with 5M km arm length and 6 laser links~\cite{Caprini:2015zlo}. }, $10^{-8} \rho_{\gamma}$~\cite{Caprini:2015zlo}.
%~\cite{Adams:2010vc,Adams:2013qma}.
%Quite model-independently from the above discussion, there is a CMB constraint on extra forms of radiation that applies to any 
%GW background, $\bar{\rho}_{\text{GW}}\lsim 0.1\rho_{\gamma}$.\footnote{This constraint usually is given in terms of the effective number of additional neutrino species, 
%$\Delta N_{eff} \lsim 0.4$ ($2 \sigma$)~\cite{Ade:2015xua,Baumann:2015rya,Brust:2017nmv}.}

% By contrast, we consider GW from first-order phase transition in very early universe.  Such phase transition is well expected, reflecting grand themes such as Grand Unified Theories, Higgs mechanisms, or Compositeness. Although the isotropic GW frequency spectrum reflects this 
%fundamental physics, the anisotropies we propose reflect even higher energy physics, including the inflation and reheating mechanisms.  The GW anisotropic components can be as large as the isotropic piece and the GW anisotropies may or may not be correlated with the CMB anisotropies.

\begin{figure}
\begin{center}
\includegraphics[height=5cm]{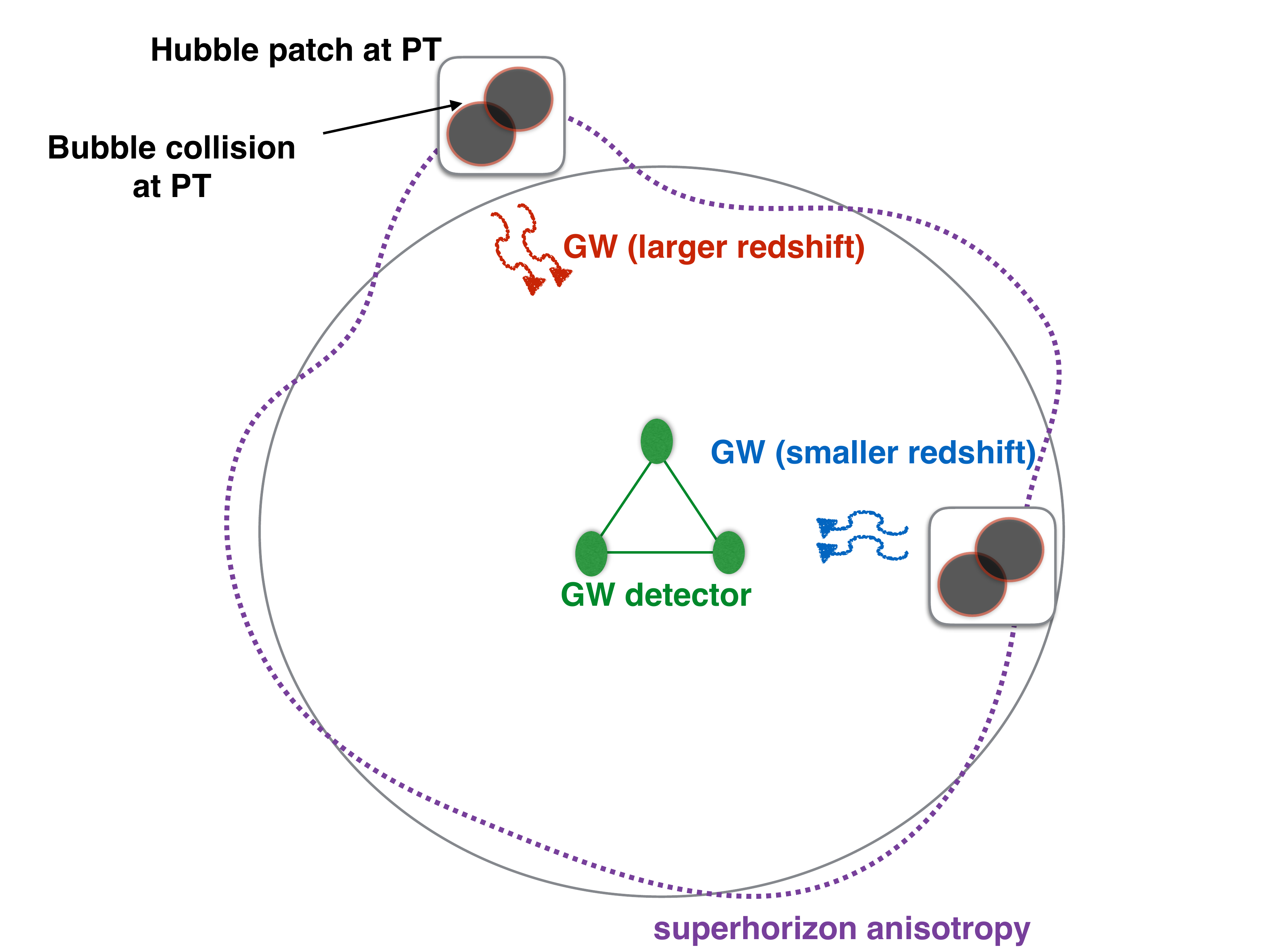}
\caption{A sketch of the anisotropies in GW produced by a cosmic PT. GW reaching the detector from different directions now, were emitted from distant regions of the universe long ago when the temperature there was $T_{PT}$, and bubbles of the low-$T$ phase were produced and collided in numerous Hubble patches. This emitting surface (dotted purple) is not a perfect sphere (solid grey) because of  temperature 
inhomogeneties. GW from more distant parts of the emitting surface will redshift longer than GW from closers parts, giving rise to  anisotropies in frequency and power. }\label{fig:sketch}
\end{center}
\end{figure}

Let us now see how anisotropies arise in the GW signals in the 
 simplest scenario when there is a single source of primordial adiabatic fluctuations.  The CMB shows us that  temperature $T$ is inhomogeneous in the universe, a redshifted reflection of inflationary quantum fluctuations. This implies that these fluctuations would also necessarily have been present during the PT
 as long as the ``reheating" temperature  at the end of inflation satisfies $T_{\text{reh}}>T_{\text{PT}}$.  Hence the PT occurs at slightly different redshifts in different patches of the sky (see Fig.~\ref{fig:sketch}), generating a non-trivial power spectrum in Eq.~(\ref{eq:ps}).  This GW anisotropy, $C_{\ell}^{\text{GW}}$, gives us a second \emph{copy} of the CMB, reflecting their shared origins. However these GW signals are created  in pristine form: the production is long before matter domination and the non-linear growth of density perturbations. Therefore, $C_{\ell}^{\text{GW}}$ is simpler and less processed than $C_{\ell}^{\text{CMB}}$  because photons interact with the charged plasma thereby feeling effects such as baryon acoustic oscillations and Silk damping at high $\ell$~\cite{Dodelson:2003ft}. 
 In principle then, $C_{\ell}^{\text{GW}}$ is proportional to the scale-invariant (SI) inflationary spectrum of fluctuations, expressed in 
 spherical harmonics. For example, {\it exact} SI would correspond to 
\begin{equation}
%{\rm SI~ Idealization:} ~ ~ ~ ~ 
C_{\ell}^{\text{SI~GW}} \propto 
[\,\ell (\ell +1)\,]^{-1},
\label{eq:scale_invariant}
\end{equation}
 but there will necessarily be small deviations directly sensitive to the physics of inflation/reheating~\cite{Baumann:2009ds}.
The only other corrections to SI  are created by the peculiar motion of the Earth, and subdominant gravitational effects of matter on GW en route to us - the ``integrated Sachs-Wolfe effect" (ISW)~\cite{1967ApJ...147...73S,Laguna:2009re}. Each of these is  however calculable.

Because we have assumed a single source of primordial fluctuations, $(\delta \rho/\rho)_{\rm GW} = (\delta \rho/\rho)_{\rm CMB} = 4.6\times10^{-5}$~\cite{Aghanim:2018eyx}. Furthermore the  two anisotropies are strongly correlated in angle: ``hot" and ``cold" regions of the CMB directly overlap ``hot" and ``cold" regions in the GW background  in the sky, 
as is captured by measuring $C^{cross}$. 
As with the CMB, the isotropic ($\ell=0$) piece of the signal will be seen first. 
Turning to the anisotropy, Eq.~(\ref{eq:rhoGW}) implies 
\begin{eqnarray}
\delta \rho_{\rm GW}&=&2.8\times 10^{-6}
\left(\frac{H_{PT}}{\beta}\right)^2\, \rho_{\gamma}, 
\end{eqnarray}
assuming $\rho_{total}=\rho_{PT}$. The range of models described by Eq.~(\ref{eq:HT}) then overlaps LISA's best sensitivity $\approx 10^{-8} \rho_{\gamma}$ at frequencies $\omega^{today}_{\rm GW}\approx1-10$ mHz and low $\ell \lsim 10$ (within LISA angular resolution~\cite{Cutler:1997ta,Kudoh:2004he}). 
 It is important that  the large isotropic  component be cleanly subtracted from the signal in making the anisotropic measurements. 
 Note that high  $\ell > 10$ modes are more challenging, both because they require higher angular resolution and because of the 
  asymptotic falloff $\sqrt{C_{\ell}^{\text{GW}}}\propto 1/\ell$ ($\ell\gg 1$), Eq.~(\ref{eq:scale_invariant}). 
   
 Even if the anisotropies are below sensitivity and only the isotropic component is detected at LISA,  
 this would specify precisely what type of future detector sensitivity, angular resolution and frequency coverage are needed to fully exploit the valuable and robustly expected anisotropies. If this turns out to be the case, there is a known astrophysical foreground \cite{Farmer:2003pa} from white dwarf mergers in the LISA frequency range and at roughly LISA sensitivity, which would become relevant for a higher-sensitivity detector. However this should  be subtractable from the signal in looking for primordial anisotropies~\cite{Cutler:2005qq,Adams:2013qma}, in part because the foreground will be dominantly from within our galaxy.

 \begin{figure}
\begin{center}
\includegraphics[height=5cm]{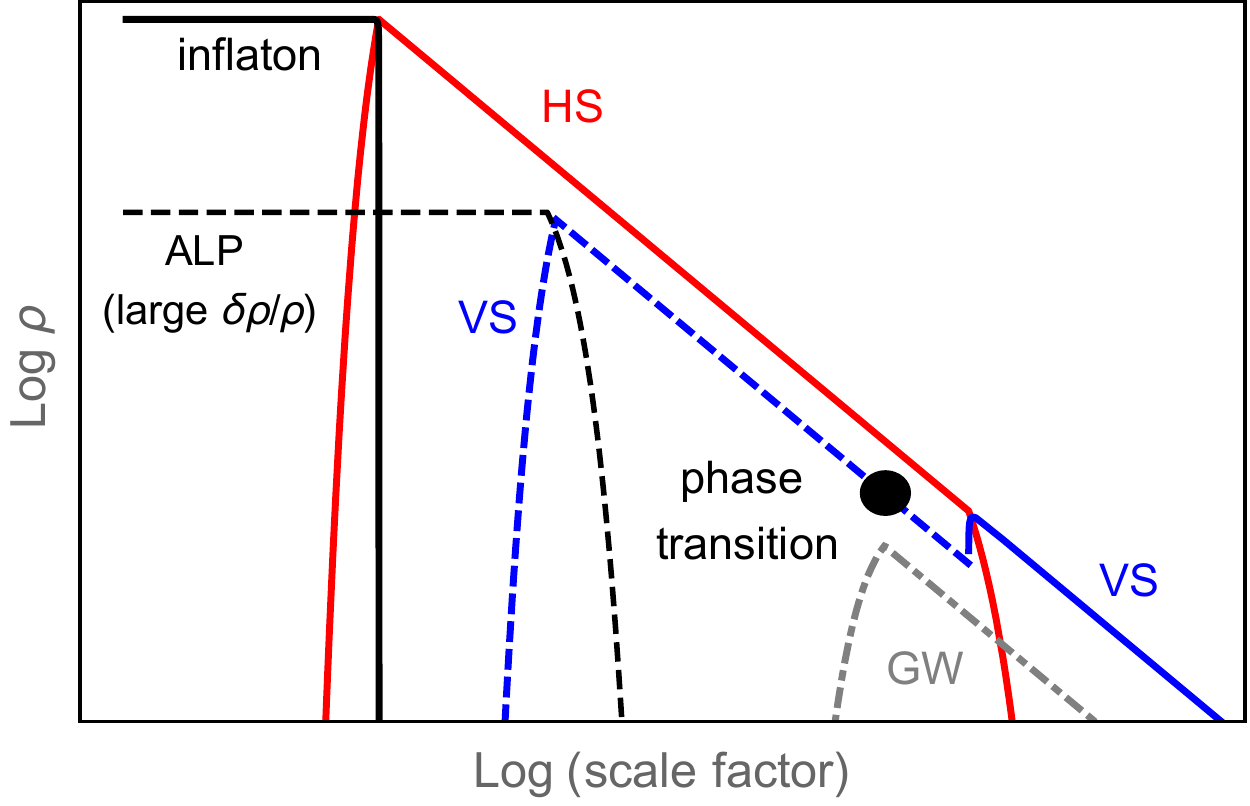}
\caption{Cosmological history of GW and CMB production and fluctuations in the presence of a hidden sector.   The inflaton (solid black) reheats the HS (red) while an ALP field (dashed black) reheats the VS (blue) in which the PT occurs. The HS particles later also decay into the VS. The dashed lines can have large contrast, 
$\delta \rho/ \rho > (\delta \rho/ \rho)_{\rm CMB} \sim 10^{-5}$, while the solid lines have $\delta \rho/ \rho \leq 10^{-5}$, saturated by the final VS (solid blue) which contains the CMB itself.  Depending on the respective sizes of fluctuations in the inflaton and the ALP, the GW signal (grey) can have different possible values of $\delta\rho/\rho$ and can be highly correlated with the CMB fluctuations or relatively uncorrelated.  }\label{fig:curvaton}
\end{center}
\end{figure}

The simple scenario above, though well motivated, is not the only option.  Observable GW anisotropies can be completely uncorrelated with the CMB, and they can also have a larger contrast $(\delta\rho/\rho)_{\rm GW} >10^{-5}$.
 The current cosmological data constrain these different options, primarily through the $N_{eff}$ and isocurvature bounds derived from the Planck satellite CMB data~\cite{Ade:2015xua}. These constraints are based on the gravitational backreaction on the CMB dynamics of the isotropic/anisotropic GW (as free-streaming dark radiation). Roughly, the isocurvature constraint is on the anisotropy, 
\begin{equation}
\delta \rho_{\rm GW} \lsim 10^{-6} \rho_\gamma,
% ~\delta \rho_{\rm GW} \lsim_{uncorrelated} 10^{-5} \rho_\gamma
\label{eq:isocurvature}
\end{equation}
while the $N_{eff}$ constraint\footnote{This constraint is usually cast in terms of the effective number of additional neutrino species, 
$\Delta N_{eff} \lsim 0.4$ ($2 \sigma$)~\cite{Ade:2015xua,Baumann:2015rya,Brust:2017nmv}.} is on the isotropic signal, 
\begin{equation}
\rho_{\rm GW} \lsim 0.1 \rho_\gamma.
\label{eq:Neff}
\end{equation}
GW satisfying these limits can easily be above LISA's sensitivity of $ 10^{-8}\rho_{\gamma}$.\footnote{Refs.~\cite{Bethke:2013vca,Bethke:2013aba} have also discussed larger contrasts appearing in high-frequency GW from inflationary preheating, but have not analyzed the cosmological constraints on this scenario.}

Detecting such GW backgrounds, either uncorrelated with the CMB and/or having a larger contrast, would be important because it would point to having two or more 
 sources of primordial fluctuations. We illustrate this with a simple example  within the inflationary paradigm, where the standard  inflaton $\phi$  is accompanied by 
  an ``axion-like particle" (ALP) (reviewed in Ref.~\cite{Marsh:2015xka}), $a$,  lighter than $H_{\rm inflation} > T_{PT}$.\footnote{An ALP is basically a pseudo Nambu-Goldstone boson associated to spontaneous breaking of an approximate $U(1)$ global symmetry.}
  Each of these fields develops approximately SI quantum fluctuations during inflation,  through repeated production, 
 $ \delta \phi, \delta a, \omega_{\phi}, \omega_{a} \sim H_{\rm inflation}$, followed by redshifting. 
  But crucially the fluctuations in the two fields are completely uncorrelated. After inflation, $\phi$ and $a$ can decay and reheat lighter particles/fields. Consider the possibility that it is the ALP that reheats some extension of the SM $+$ DM, 
  which we will call the ``visible sector" (VS), while the inflaton reheats another 
   ``hidden" sector (HS) of light particles which only couples to VS very weakly. Therefore, the VS inherits the quantum fluctuations of $a$, and the HS inherits the quantum fluctuations of $\phi$, so that the fluctuations in the two sectors are uncorrelated. 
   Because the inflaton, by definition, carries  the most (potential) energy until the end of inflation, it can naturally dominate the reheating process, 
   so that $\rho_{\rm HS} > \rho_{\rm VS}$, with the parameters of the inflaton/ALP dynamics and decay allowing a very large range of ratios $\rho_{\rm VS}/\rho_{\rm HS}$.

 Subsequently, the VS and HS redshift and cool, until the PT is reached in the VS and GW are released.
  At some time later, only bounded to be earlier than Big Bang Nucleosynthesis  (as reviewed in Ref.~\cite{Fields:2006ga}), $T \sim$ MeV, we take the HS particles to decay entirely into the VS, via very weak couplings, leaving just  the VS in thermal equilibrium.  
  This cosmological history is illustrated in Fig.~\ref{fig:curvaton}. 
 Late-decaying HS decays to the VS have been been discussed earlier as an attractive setting for generating the matter/antimatter asymmetry in baryons~\cite{Davidson:2008bu,McDonald:2011zza,Cui:2012jh} as well as for some variants of DM~\cite{Feng:2003uy,Pospelov:2008jk,Zurek:2013wia}.
   
We now turn to the history of the primordial fluctuations. Standard single-field inflationary dynamics readily accommodates small fluctuations imparted  by $\phi$ decay to the HS, say compared to the CMB,
\begin{equation}
\left(\frac{\delta \rho}{\rho}\right)_{\rm HS} \leq  4.6 \times 10^{-5}.
\end{equation}
However, the ALP can easily give rise to larger contrast, as follows.  The field space of an ALP Goldstone boson is compact,  its size characterized by its  ``decay constant", $f_a$. If $f_a > H_{\rm inflation}$, so that the associated spontaneous symmetry breaking has already taken place during inflation, then it is
natural for $a$ to have a background value of order $f_a$, with quantum fluctuations $\delta a \sim H_{\rm inflation}$. This leads to 
\begin{equation}
\left(\frac{\delta \rho}{\rho}\right)_{{\rm VS}} \sim \frac{\delta a}{a} \sim \frac{H_{\rm inflation}}{f_a}   \geq  5 \times 10^{-5}
\end{equation} 
being satisfied for a broad range of parameters.  Once the VS undergoes its PT, it imparts its $\delta \rho/\rho$ to its 
 low-$T$ phase as well as the GW,
 \begin{equation} 
\left(\frac{ \delta \rho}{\rho}\right)_{\rm GW} = \left( \frac{\delta \rho}{\rho}\right)_{\rm VS}  \geq 5 \times 10^{-5}.
\end{equation}
But after HS decay to VS, we have
\begin{equation}
(\delta) \rho_{\rm VS}^{\rm after} = (\delta) \rho_{\rm VS}^{\rm before} +  (\delta) \rho_{\rm HS}^{\rm before}, 
\end{equation}
while $(\delta) \rho_{\rm GW}$ is unaffected, 
implying that today the CMB has
\begin{eqnarray}\label{eq:drhoCMB}
&&\left(\frac{\delta \rho}{\rho}\right)_{\rm CMB}  \approx \frac{ \delta \rho_{\rm VS}^{\rm after}}{   \rho^{\rm before}_{\rm HS}}\nonumber\\
&& ~ ~ =\left (\frac{\rho_{\rm VS}}{\rho_{\rm HS}}\right)^{\rm before} \left( \frac{\delta \rho}{\rho}\right)_{\rm GW} + \left( \frac{\delta \rho}{\rho}\right)^{\rm before}_{\rm HS}.
\end{eqnarray}
Here, we have used that $\rho_{\rm HS} > \rho_{\rm VS}$ originally. We see that we can easily reconcile $(\delta \rho/\rho)_{\rm CMB} = 4.6 \times 10^{-5}$ with {\it larger} $
( \delta \rho/\rho)_{\rm GW}$ by having $\rho_{\rm HS} > \rho_{\rm VS}$ originally. Furthermore, we see two different patterns depending on whether the first or second term on the second line dominates: if the first term dominates then the CMB and GW backgrounds are completely correlated, even if $(\delta \rho/\rho)_{\rm GW} > (\delta \rho/\rho)_{\rm CMB} $, 
while if the second term dominates then the CMB and GW backgrounds are completely uncorrelated. For instance, in the uncorrelated case, if the effects of ISW were removed from the data, then the cross-correlator would vanish, $C^{cross} = 0$.

The separate evolution of the fluctuations in the VS and HS clearly requires that the two sectors are substantially decoupled. But minimally they must interact via gravity, as well as by the requirement that the HS decays into the VS after the PT. One danger is that large fluctuations in the VS, $\delta \rho_{\rm VS}$, can 
source spacetime curvature fluctuations,  in turn adding to $\delta \rho_{\rm HS}$ beyond that inherited from the inflaton, threatening the ability to reconcile the small CMB anisotropy with large GW anisotropy in Eq.~(\ref{eq:drhoCMB}).
%\footnote{Although gravity is a  very weak force, it can have order one effects acting over cosmological periods of order $1/H$.} 
This danger is  simply avoided by requiring that $\delta \rho_{\rm VS} \leq \delta \rho_{\rm HS}$ originally, even though 
$(\delta \rho/\rho)_{\rm VS} \geq (\delta \rho/\rho)_{\rm HS}$. The second danger is HS-VS interactions may equilibrate the two sectors before the PT.
But the interaction rate per HS particle can naturally be as small as the decay rate,  which is $< H_{PT}$, corresponding to decays after the PT. 
This then ensures that the interactions are also ineffective in equilibrating before the PT.

Note that the above mechanism for generating large  $( \delta \rho/\rho)_{\rm GW} > 10^{-5}$ implies that the GW signal strength is weaker than the minimal scenario of a single source of primordial fluctuations, first discussed.  This can be seen from Eq.~(\ref{eq:rhoGW}), which translates here to 
\begin{eqnarray}
\rho_{\rm GW} 
&\approx& \left(\frac{\rho_{\rm VS}}{\rho_{\rm HS}}\right)^2 \rho_{\rm GW}^{\rm single\mbox{-}source}.
\end{eqnarray}
With the weaker signal, the $N_{eff}$ and isocurvature constraints, Eqs.~(\ref{eq:isocurvature}, \ref{eq:Neff}), are automatically satisfied. There is also an extra relative GW redshift, due to the final large influx of energy from HS decay to the VS but not to GW,  
\begin{eqnarray}\label{eq:ftoday}
\omega_{\rm GW, today} &\approx&  \left ( \frac{\rho_{\rm HS}}{\rho_{\rm VS}} \right )^{1/4} \omega_{\rm GW, today}^{\rm single\mbox{-}source}.
\end{eqnarray}

This GW signal can still be above future detector sensitivity.
For example, $\rho_{\rm VS}  = 0.5 \rho_{\rm HS}$ would allow
$(\delta \rho/\rho)_{\rm GW}$ to be comparable in magnitude with the CMB and yet only partially correlated with it. 
For a PT with $ \beta^2 = 10\,H_{PT}^2$, this gives an anisotropic signal, 
$\delta \rho_{\rm GW} \approx 10^{-7} \rho_{\gamma}$, above LISA sensitivity up to $\ell \sim$ few, with $\omega_{\rm GW} \approx 1$ mHz for $T_{PT}=10$ TeV.
Alternatively, we can have larger GW anisotropy $(\delta \rho/\rho)_{\rm GW} \approx 5 \times 10^{-4}$, either {\it correlated or uncorrelated} with the CMB, for $\rho_{\rm HS} = 10 \rho_{\rm VS}$.   For a PT with $ \beta^2 = 100\,H_{PT}^2$, this gives an anisotropic signal, $\delta \rho_{\rm GW} \approx 3 \times 10^{-9} \rho_{\gamma}$, above the sensitivity of the proposed Big Bang Observatory (BBO)~\cite{Crowder:2005aa} of $10^{-11} \rho_{\gamma}$ 
(up to $\ell \sim 100$, within BBO angular resolution~\cite{Cutler:2009qv}), with $\omega_{\rm GW} \approx 0.1$ Hz for $T_{PT}\approx 200$ TeV. 
An even weaker signal with $ \beta^2 = 10^{4}H_{PT}^2$ can still have fluctuations $(\delta \rho/\rho)_{\rm GW}$ comparable to the CMB and yet uncorrelated with it, with 
$\delta \rho_{\rm GW} \approx 10^{-10} \rho_{\gamma}$, 
still visible at BBO up to $\ell \sim 10$. Again, there are 
astrophysical merger foregrounds relevant to LISA and BBO \cite{Farmer:2003pa} but these should be predominantly isotropic or subtractable~\cite{Adams:2013qma}. But by frequencies $\omega_{\rm GW} \gsim$ Hz (for sufficiently high $T_{PT}$), these foregrounds are essentially absent. 

We have shown that stochastic GW from PT in extensions of the SM can have observable anisotropies, giving valuable new insights into inflation and reheating. Similarly,  PT in completely hidden sectors~\cite{Schwaller:2015tja}  can also give GW anisotropies. Beyond the two-point correlators discussed above, it might eventually be possible to detect three-point correlators, giving further information on inflationary dynamics.

~\\
\mysection{Acknowledgments}
We are very grateful to Nima Arkani-Hamed, Alessandra Buonanno, Yanou Cui, Gulia Cusin, Kaustubh Deshpande, Soubhik Kumar, Marc Kamionkowski, Surjeet Rajendran, 
Fabian Schmidt, Pedro Schwaller, Peter Shawhan, and Matias Zaldarriaga for useful discussions.  This research was supported in part by the NSF under Grant No. PHY-1620074 and by the Maryland Center for Fundamental Physics (MCFP).
%\bibliography{ref}

\bibliography{AGW}

\end{document}